\documentclass{INTERSPEECH2023}

\interspeechcameraready

\usepackage[title]{appendix}
\usepackage{amsmath,amsfonts}
\usepackage{amssymb}
\usepackage{mathtools}
\usepackage{algorithm}
\usepackage{algorithmic}
\usepackage[linesnumbered,ruled, algo2e]{algorithm2e}
\usepackage{setspace, fullwidth}
\usepackage{lipsum}

\usepackage{array}
\usepackage[caption=false,font=normalsize,labelfont=sf,textfont=sf]{subfig}
\usepackage{textcomp}
\usepackage{stfloats}
\usepackage{url}
\usepackage{verbatim}
\usepackage{graphicx}
\usepackage{cite}
\usepackage{orcidlink}

\usepackage{multirow}
\usepackage{multicol}

\usepackage{booktabs}
\usepackage{tcolorbox}
\usepackage{flushend}
\usepackage{makecell}
\usepackage{colortbl}
\usepackage{color}
\usepackage{threeparttable} % to use table notes

\usepackage{dsfont}
\usepackage{amsmath,amssymb}

\usepackage{comment}

\newcolumntype{C}[1]{>{\PreserveBackslash\centering}p{#1}}

\title{Range-Based Equal Error Rate for Spoof Localization}
\name{Lin Zhang$^{1,2}$, Xin Wang$^1$, Erica Cooper$^1$, Nicholas Evans$^3$, Junichi Yamagishi$^{1,2}$}
\address{
  $^1$National Institute of Informatics, Tokyo, Japan \\
  $^2$SOKENDAI (The Graduate University for Advanced Studies), Kanagawa, Japan\\
  $^3$Digital Security Department, EURECOM, France}
\email{ \{zhanglin, wangxin, ecooper, jyamagis\}@nii.ac.jp, evans@eurecom.fr }

\begin{document}

\maketitle
 
\begin{abstract}
Spoof localization, also called segment-level detection, is a crucial task that aims to locate spoofs in partially spoofed audio. The equal error rate (EER) is widely used to measure performance for such biometric scenarios. Although EER is the only threshold-free metric, it is usually calculated in a point-based way that uses scores and references with a pre-defined temporal resolution and counts the number of misclassified segments. Such point-based measurement overly relies on this resolution and may not accurately measure misclassified ranges. To properly measure misclassified ranges and better evaluate spoof localization performance, we upgrade point-based EER to range-based EER. Then, we adapt the binary search algorithm for calculating range-based EER and compare it with the classical point-based EER. Our analyses suggest utilizing either range-based EER, or point-based EER with a proper temporal resolution can fairly and properly evaluate the performance of spoof localization.

% Our implementation is based on pyannote and will be publicly available after the paper's publication.

\end{abstract}

\section{Introduction}
\label{sec:intro}

Automatic speaker verification (ASV) is vulnerable to spoofing attacks (also known as presentation attacks or PA) \cite{isoiec2016}. Some challenges were thus held to encourage the development of countermeasures (CMs) to protect ASV from spoofing, such as ASVspoof \cite{Wu2014, Kinnunen2017, Nautsch2021spoof19, Wang2020data, asvspoof2021, liu2022asvspoof} and ADD \cite{Yi2022ADD}. CMs for those challenges operate at the utterance level to detect whether an utterance is spoofed. The equal error rate (EER) and tandem detection cost function (t-DCF) \cite{Kinnunen-tdcf} are then commonly used to evaluate the performance of CMs and consistently measure the progress in this field over time.

Partial Spoof (PS) \cite{zhang2022partialspoof} is a recently proposed spoofing scenario in which only a fraction of speech utterances are spoofed. It is one of the most important and challenging scenarios for the anti-spoofing community as detecting a fraction of speech segments is much more difficult than detecting a whole spoofed utterance. Accordingly, besides conventional utterance-level detection, spoof localization, also known as segment-level detection \cite{zhang21partialspoof_mtl, Zhang2021PartialSpoof, Yi2021halftruth}, was designed for the PS scenario. Spoof localization aims to locate spoofed regions within partially spoofed audio, that is, to answer \textit{``when do spoofs happen?''}. Spoof localization is an important task for the PS scenario that can be used as a pre-processing step and provides cues to further analyze attackers' intentions.

It is also crucial to evaluate different models for spoof localization to make progress. However, as a newly introduced task in the anti-spoofing community, there is currently no established way of properly measuring the performance of spoof localization. Most metrics usually face dilemmas \cite{nips2021goodmetric} and depend on a pre-defined threshold. Furthermore, the use of different measurements, such as counting the number of misclassified segments with fixed temporal resolutions (10 ms \cite{zhang2022localizing}, 20 ms \cite{zhang2022partialspoof}) or measuring the duration of misclassified regions \cite{Yi2021halftruth}, hinders the comparison of different spoof localization methods across the literature. Following \cite{tatbul2018nipsmetric}, we named the former approach of counting the number of misclassified segments ``point-based'' measurement and the latter approach of measuring the duration of misclassified regions ``range-based'' measurement.

For example, Yi \MakeLowercase{\textit{et al.}}~\cite{Yi2021halftruth} used range-based precision, recall, and F1 to measure performance in accurately detecting spoofed regions. However, these require a pre-defined threshold and have a high bias on imbalanced data \cite{LUQUE2019imbalancemetric}. 
Furthermore, Zhang \MakeLowercase{\textit{et al.}}~\cite{zhang2022localizing} utilized point-based IoU (intersection over union, also known as the Jaccard index) by counting the number of accurately predicted frames to describe the similarity between reference and prediction. All of the above metrics depend on a pre-defined threshold. In contrast, we ~\cite{Zhang2021PartialSpoof} adapted a threshold-free EER from the utterance level to the segment level. However, it is still a point-based EER that requires a pre-defined temporal resolution for reference and it is easy to ignore some misclassified regions that can only be measured at high precision.
% something with a high degree of accuracy during the evaluation. 
Range-based EER is a possible solution to measuring such misclassified regions properly.
% However, it usually requires iterative searching of mismatched regions in each trial based on all possible thresholds, which can be computationally expensive.

% Thus,
% it is necessary to propose an efficient way to calculate range-based EER. Therefore, 
To estimate range-based EER for measuring the performance of spoof localization in the PS scenario, we adapted the binary search algorithm (also known as the half-interval search method \cite{binarysearch1976}). Then, we compared range-based EER with point-based EER for better understanding. 

Our results show that when the temporal resolution of a point-based reference is coarser than the temporal resolution of the training data, point-based evaluation becomes too coarse. For fair and proper evaluation of spoof localization models, we recommend using range-based EER, or point-based EER that uses references with a finer temporal resolution than that used to train spoof localization models.

% In addition, the range-based EER has the potential to be utilized for evaluating other binary classification tasks in the time series and promote long-term technical progress.

The remainder of this paper is structured as follows. Section \ref{sec:point} introduces the basic properties of point-based measurement and its relationship with range-based measurement. Section \ref{sec:range} describes range-based EER and proposes the adapted binary search algorithm to estimate range-based EER for spoof localization. Section \ref{sec:exp} introduces the experiments and discusses the relationship between point-based and range-based EER. Finally, Section \ref{sec:conclusion} gives the conclusion.

\section{Point-Based Measurement}
\label{sec:point}
In this section, we first introduce the widely used point-based measurement. Then, we extend point-based measurement to range-based measurement, providing a foundation for understanding the basic properties of range-based EER, which we will discuss in the next section.

\subsection{Two types of errors: false positive and false negative}
\label{sec:two_errors}
Following the ISO/IEC standard~\cite{isoiec2016}, we treat \textit{spoof} as positive and \textit{bona fide} as negative. Then, CMs are subject to two types of errors: false positive and false negative\footnote{Note that FA and MD in \cite{Wu2014}, and FR and FA in \cite{wang2022chapter} are equivalent to FP and FN in this paper, respectively. Their names are different, but they share the same definition for the spoof scenario.}:

\begin{itemize}
    \item FP (False Positive): the number or duration of \textit{bona fide} misclassified as \textit{spoof}.
    \item FN (False Negative): the number or duration of \textit{spoof} misclassified as \textit{bona fide}.
\end{itemize}

The normalized (proportional) versions of FP and FN are called the false positive rate (FPR) and false negative rate (FNR). The corresponding confusion matrix is shown in Table \ref{tab:conf_matrix},
% \footnote{Based on this confusion matrix, we have re-formula existing metrics in spoof localization. You can find them in the appendix of the arXiv version.}
where TP (true positive) and TN (true negative) refer to correctly predicted \textit{spoof} and \textit{bona fide}, respectively. They can be calculated by either counting the number of segments or measuring the duration of eligible regions.
% which will be introduced next. 

\begin{table}[!t]
\caption{Confusion matrix for spoof localization.}
\label{tab:conf_matrix}
\renewcommand\arraystretch{1.35}
\centering
\begin{threeparttable}
% \begin{tabular}{cc|cc}
\setlength\tabcolsep{0pt}
\begin{tabular*}{\linewidth}{@{\extracolsep{\fill}} cc|cc }

\toprule
                              &                             & \multicolumn{2}{c}{Hypothesis}                                                \\
                              &                             & \textbf{Positive (\textit{spoof})} & \textbf{Negative (\textit{bona fide})}                       \\
                              \midrule
\multirow{2}{*}{\rotatebox{90}{\footnotesize Reference}} & \textbf{Positive} ($\mathcal{P}$)   & $TP$                        & {$FN$}                      \\
                              & \textbf{Negative} ($\mathcal{N}$) & {$FP$}                & $TN$        \\
          \bottomrule
% \end{tabular}
\end{tabular*}
    \end{threeparttable}
\vspace{-5mm}
\end{table}

\subsection{Classical point-based EER}
Point-based EER is widely utilized for the binary classification task. It is a threshold-free metric and is the error rate with a specific threshold where the FPR is closest to the FNR.
% And point-based EER is commonly computed. 
Following the predicted scores and confusion matrix defined in Table \ref{tab:conf_matrix}, we express the definition of FPR and FNR as follows:

\begin{equation}
    \label{eq:point_fpr}
    \begin{aligned}
    P_\text{FP}(\tau) &= \frac{1}{| \varLambda^{p}_{\mathcal{N}}| } \sum_{\phantom{0}m \in \varLambda^{p}_{\mathcal{N}} } \mathds{1}(s_m < \tau), \\
    \end{aligned}
\end{equation}
\begin{equation}
    \label{eq:point_fnr}
    \begin{aligned}
    P_\text{FN}(\tau) &= \frac{1}{| \varLambda^{p}_{\mathcal{P}}| } \sum_{\phantom{0}m \in \varLambda^{p}_{\mathcal{P}} } \mathds{1}(s_m \geq \tau), \\
    \end{aligned}
\end{equation}
where both $P_\text{FP}(\tau)$ and $P_\text{FN}(\tau)$ are functions of a pre-defined threshold $\tau$. $\varLambda^{p}_{\mathcal{N}}$ and $\varLambda^{p}_{\mathcal{P}}$ index bona fide and spoof \textit{segments}, respectively. Then, $|\varLambda^{p}_{\mathcal{N}}|$ and $|\varLambda^{p}_{\mathcal{P}}|$ denote the total number of bona fide and spoof segments separately. $s_{m}$ is the segment score for the $m$-th segment as shown in Fig.~\ref{fig:seg_dur}(a). $\mathds{1}(\cdot)$ denotes the indicator function that outputs 1 when the condition is true and 0 otherwise.

\begin{comment}
\begin{equation}
\label{eq:point_fpr}
    P_\text{FP}(\tau) = \frac{1}{|\textcolor{red}{\mathcal{N}}|}\sum_{\phantom{0}x_{m} \in \textcolor{red}{\mathcal{N}}} \mathds{1}(s_{m} < \tau),
\end{equation}
\begin{equation}
\label{eq:point_fnr}
    P_\text{FN}(\tau) = \frac{1}{|\textcolor{red}{\mathcal{P}}|}\sum_{\phantom{0}x_{m} \in \textcolor{red}{\mathcal{P}}} \mathds{1}(s_{m} \geq \tau), \\
\end{equation}
where $\mathcal{P}$ and $\mathcal{N}$ are the segment-level data sets for positive (spoof) and negative (bonafide), respectively. 
% And $\mathcal{D} = \{\mathcal{P} \cup \mathcal{N}\}$ is the data set. 
$s_{m}$ is the segment score for the $m$-th segment as shown in Fig.~\ref{fig:seg_dur}(a). $\tau$ presents a pre-defined threshold, and both $P_\text{FP}(\tau)$ as well as $P_\text{FN}(\tau)$ are functions of $\tau$. $\mathds{1}(\cdot)$ denotes the indicator function that outputs 1 when the condition is true and 0 otherwise.
\end{comment}

EER is decided by $\hat\tau$ where the value of $P_\text{FP} (\hat\tau)$ is infinitesimally close to $P_\text{FN} (\hat\tau)$. Then, EER can be computed by:
\begin{equation}
\label{eq:eer}
    EER = \frac{P_\text{FP} (\hat\tau)  + P_\text{FN} (\hat\tau) }{2}, \\
\end{equation}
where
\begin{equation}
\label{eq:threshold}
    \hat\tau = \arg \min_{\tau} |P_\text{FP} (\tau) - P_\text{FN} (\tau) |.
\end{equation}
% where $\tau$ is a pre-defined threshold. And $EER$ is decided by $\hat\tau$ where the value of $FPR$ is infinitely close to $FNR$. When $\tau = \hat\tau$, half the total error rate ($HTER$) is equal to $EER$, where $HTER = \frac{FPR + FNR}{2}$. This relationship between $HTER$ and $EER$ is widely used to estimate $EER$.

\subsection{From point-based to range-based measurement}
\label{sec:point2range}
\begin{figure}[!t]
\centerline{\includegraphics[width=1\columnwidth]{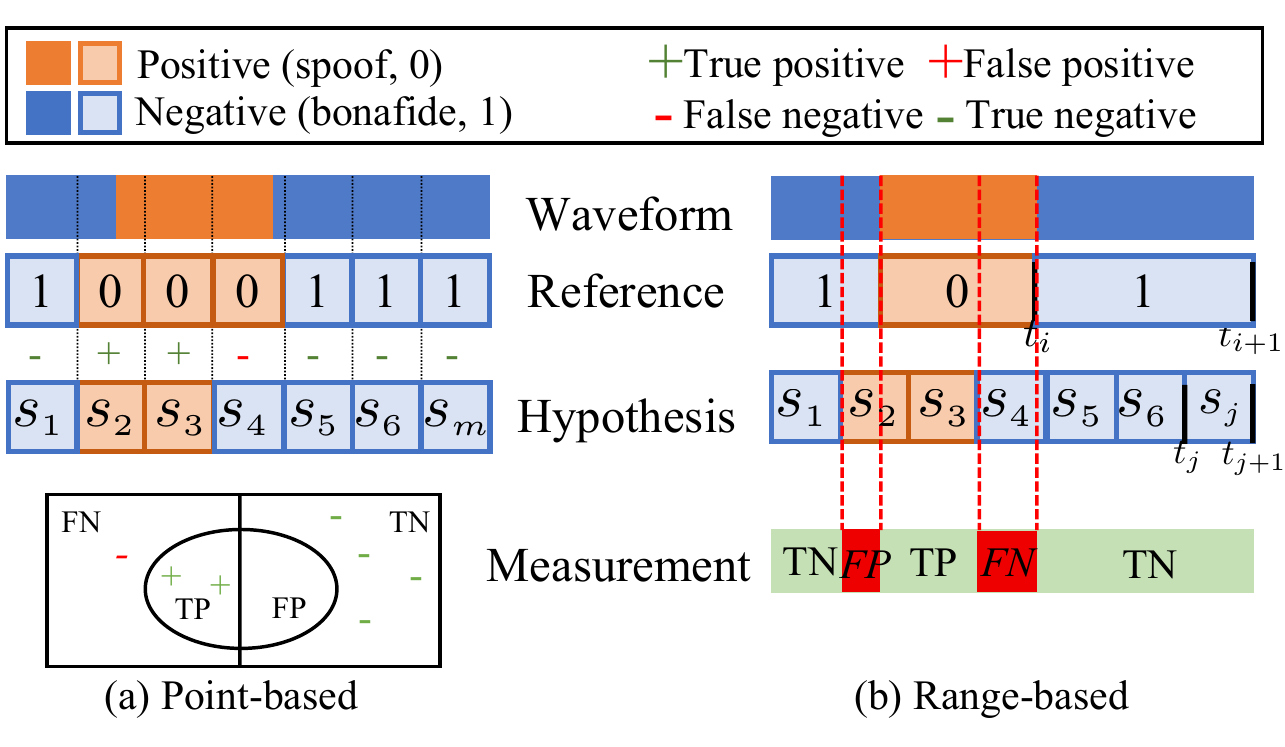}}
\vspace{-2mm}
\caption{Comparison of point-based and range-based measurement. (In the Reference row of this example, there are 7 uniform points in (a) and 3 varied ranges in (b))}
\label{fig:seg_dur}
\vspace{-8mm}
\end{figure}

Although the common evaluation method is to utilize point-based measurement, it can be implemented by range-based measurement for time series problems.
Depending on the method of measuring the predicted results, all properties of the previously defined confusion matrix can be calculated using either the point-based or range-based\footnote{Those two levels are called ``classical'' and ``durative'' in \cite{modaresi2022uniform}, and ``frame-based'' as well as ``boundary-based'' in \cite{tong2014evaluating}.} measurements as illustrated in Fig. \ref{fig:seg_dur}.

As for the point-based measurement shown in Fig.~\ref{fig:seg_dur} (a), we need to split the audio into uniform segments with a fixed resolution and assign corresponding reference labels. Then, we measure the performance on the basis of a comparison between those pre-segmented labels and discrete predicted segment scores. However, this point-based measurement can be easily influenced by the resolution of the references. This is because each uniform segment corresponds to only one label, and the segment is more likely to contain different classes with coarser resolution. This could result in imprecise evaluation results. Besides, a finer resolution can ensure that the segment is more likely to have only one class, thereby improving the precision of the evaluation. 
% \textcolor{red}{Our experiments in Section \ref{sec:exp} will prove that a resolution finer than the training one is more appropriate.}

% Thus, point-based measurement can be easily influenced by the resolution of references.
% Therefore, point-based measuring usually supports directly evaluating the output in the uniform unit but not time-varying output. 
% In addition, point-based measurement requires recording scores for all segments, which incurs a high space complexity.
% but offers faster evaluation.

In contrast, we do not need to do pre-segmentation for the range-based measurement as shown in Fig.~\ref{fig:seg_dur} (b). Instead, we need to measure the duration of misclassified regions between references and hypotheses of each trial with higher precision. Thus, range-based measurement needs to record the boundaries for bona fide or spoof regions in the references and hypotheses. But it does not require a definition of resolution.
% , but the pre-processing to measure \textcolor{red}{mismatched duration between reference and hypothesis} for each trial is \textcolor{red}{computationally expensive}.

\section{Range-Based Measurement}
\label{sec:range}

% As discussed in section \ref{sec:intro}, EER is the only threshold-free metric. Following section \ref{sec:two_levels}, metrics at a continuous level can measure performance with high precision. Therefore, $conEER$ would be a good choice for spoof localization. But computing is time-consuming with finding mismatches for each trial based on all thresholds. Thus, we adapted the binary search algorithm algorithm to estimate $conEER$.

% In this section, we first introduced the classical discrete-based EER. Then, we introduced binary search algorithm and its adaption for efficiently estimating $conEER$.

\subsection{Range-based FPR, FNR, and EER}

Although the implementation of measurements in existing literature may be different\footnote{Different measurements are utilized in the PS scenario. Point-based measurement: EER in \cite{Zhang2021PartialSpoof} and IoU in \cite{zhang2022localizing}. Range-based measurement: precision, recall, and F1 in \cite{Yi2021halftruth}.}, all existing metrics are defined on the basis of the confusion matrix in Table \ref{tab:conf_matrix} and can be calculated by using point-based or range-based measurement. 
% \textcolor{red}{Given a threshold, we can derive the hypothesis as shown in Fig.~\ref{fig:seg_dur} (b).}
Suppose a hypothesis is given by a segment-level detection model and each segment in the hypothesis has a score as shown in Fig.~\ref{fig:seg_dur}(b). 
Then, the range-based version of Eqs. (\ref{eq:point_fpr}) and (\ref{eq:point_fnr}) can be formulated as follows:
\begin{equation}
    \begin{split}
    \label{eq:range_fpr}
            P_\text{FP}(\tau) =  \frac{1}{D_\mathcal{N} } \sum_{i \in \varLambda^{r}_{\mathcal{N}} } \sum_{j} \mathds{1}(s_j < \tau)\mathds{T}(r_i, r_j) ,
    \end{split}
\end{equation}
\begin{equation}
    \begin{split}
    \label{eq:range_fnr}
            P_\text{FN}(\tau) =  \frac{1}{D_\mathcal{P} } \sum_{i \in \varLambda^{r}_{\mathcal{P}} } \sum_{j} \mathds{1}(s_j \geq \tau)\mathds{T}(r_i, r_j),
    \end{split}
\end{equation}
where $i$ and $j$ index the time range in the reference and hypothesis, separately.  $\varLambda^{r}_{\mathcal{N}}$ and $\varLambda^{r}_{\mathcal{P}}$ index bona fide and spoof \textit{ranges} in references. $s_j$ is the predicted score derived from the CM for the $j$-th range\footnote{The duration of the $j$-th range in the hypothesis can either be uniform with a resolution of $d = t_{j+1} - t_{j}$ or a variable length of $t_{j+1} - t_{j}$. Experiments in Section \ref{sec:exp} of this paper belong to the former case.}. $D_\mathcal{N}$ and $D_\mathcal{P}$ respectively denote the total duration of bona fide and spoof \textit{ranges}, where $D_\mathcal{N} = \sum_{i \in \varLambda^{r}_{\mathcal{N}}}\mathds{T}(r_i, r_i) $ and $D_\mathcal{P} = \sum_{i \in \varLambda^{r}_{\mathcal{P}}}\mathds{T}(r_i, r_i)$. $\mathds{T}(r_i, r_j)$ denote the overlapped duration between two ranges $r_i$ and $r_j$. $r_i$ is the range with the start time $t_i$ and end time $t_{i+1}$. Then, $\mathds{T}(r_i, r_j)$ can be formulated as:
\begin{equation}
\mathds{T}(r_i, r_j) = \max(0, \min(t_{i+1}, t_{j+1}) - \max(t_i, t_j)).
\end{equation}

However, calculating EER usually requires comparing $P_\text{FP}(\tau)$ and $P_\text{FN}(\tau)$ on the basis of all possible $\tau$. Thus, we adapted the binary search algorithm to find $\hat\tau$ and estimate range-based EER.

\subsection{Binary search algorithm for  range-based EER}
The binary search algorithm is a method that efficiently searches for a value in a sorted list of elements. The algorithm works by dividing the list in half at each iteration and determining which half of the list the target value is in. We adapted the binary search algorithm to estimate range-based EER as shown in Algorithm \ref{al:bisec_conEER}\footnote{https://github.com/nii-yamagishilab/PartialSpoof}. The notations used are shown in Table \ref{tab:notations}. Subscripts $l, m$, and $r$ represent the left, middle, and right values respectively in the region of each iteration during binary search. To better adapt the binary search algorithm and estimate range-based EER, we made two modifications:

\begin{enumerate}
    \item We divided the list in half on the basis of the quantile but not the value as shown in line 13 of Algorithm \ref{al:bisec_conEER}. Given that the EER is calculated by score distribution, we searched $\hat\tau$ on the basis of the \textit{quantile} ($\frac{Q_l+Q_r}{2}$) but not the value ($\frac{\tau_l + \tau_r}{2}$) as in the original binary search algorithm.

    \item We introduced an additional condition $(P_\text{FP}(\tau_l) - P_\text{FN}(\tau_l)) \times (P_\text{FP}(\tau_m) - P_\text{FN}(\tau_m)) \le 0$ to assign the value for the middle threshold as shown in line 8. When we have a predicted score that refers to how likely it would be bona fide, $P_\text{FP}(\tau)$ is an increasing function while $P_\text{FN}(\tau)$ is a decreasing function of the threshold $\tau$. Here is an important theorem, that is, when $\tau < \hat\tau $, $P_\text{FP}(\tau) < P_\text{FN}(\tau)$ and vice versa \cite{poh2006database}. Thus, in addition to the common condition in the binary search algorithm, we utilized $(P_\text{FP}(\tau_l) - P_\text{FN}(\tau_l)) \times (P_\text{FP}(\tau_m) - P_\text{FN}(\tau_m)) \le 0$ to assign the value for the middle threshold $\tau_m$.    
\end{enumerate}

\section{Experiments}
\label{sec:exp}
 To further explore the relationship between point-based and range-based measurement, we measured EER on a recent powerful model \cite{zhang2022partialspoof} for spoof localization. This section introduces the database and experimental configuration.

\begin{table}[htb]
    \centering
    \caption{Notations used in the binary search algorithm.}
    % \caption{List of math notations used in this paper.}
    \label{tab:notations}
    \begin{tabular}{rl}
    \toprule
    $\boldmath{s}$& List of sorted predicted scores, \\
    $\boldmath{y}$& List of ground-truth labels, \\
    $\tau, Q_\tau$& Threshold and its quantile, $Q_\tau \in [0,100]$, \\
    $\tau_l, Q_l$& Lower threshold and its quantile, \\ 
    $\tau_r, Q_r$& Upper threshold and its quantile, \\
    $\tau_m, Q_m$& Middle threshold and its quantile, \\
    $prec$& Precision we want to get. \\
    \bottomrule
    \end{tabular}
    \vspace{-5mm}
\end{table}

\begin{fullwidth}[width=1\linewidth,leftmargin=0.2cm,rightmargin=0cm]
\begin{algorithm}[H]
% \setstretch{1.25}
\caption{Binary search algorithm for range-based EER.}
\label{al:bisec_conEER}
\SetAlgoLined
\DontPrintSemicolon
\SetKwInOut{Input}{Input}
\SetKwInOut{Output}{Output}
\Input{ $\boldmath{s}, \boldmath{y}, prec$
% \texttt{// $prec = 1e-5$ in this paper.}
}

\Output{Estimated range-based $EER$}
\BlankLine
  \DontPrintSemicolon
  \SetKwFunction{FPRFNR}{Cal\_FPR\_FNR}
  \SetKwProg{Fn}{Function}{:}{\KwRet {$P_\text{FP}(\tau)$, $P_\text{FN}(\tau)$}}
  \Fn{\FPRFNR{$\boldmath{s}$, $\boldmath{y}$, $\tau$ }}{
  \texttt{
  //Utilize Eqs. (\ref{eq:range_fpr}) and (\ref{eq:range_fnr}) to calculate range-based FPR and FNR on the basis of the threshold $\tau$.} \\
  } 
 
  \SetKwFunction{Percentile}{Percentile}
  \SetKwProg{Fn}{Function}{:}{\KwRet {percentile value}}
  \Fn{\Percentile{$\boldmath{s}$, $Q$ }}{
    \texttt{//Get percentile value of $Q$ in $\boldmath{s}$.}
  }
  
  % \SetKwFunction{Initial}{Initial}
  % \SetKwProg{Fn}{Function}{:}{}
  % \Fn{\Initial{$scores$, $labels$}}{
  %   $left = min(scores)$ \\
  %   $mid = \Cal_EER(scores, labels)$ \\
  %   $right = max(scores)$ \\
    
  %   \KwRet $left, mid, right$ \;
  % }
  % \;
  
\While{$\tau_l \leq \tau_r$ AND $abs(P_\text{FP}(\tau_m) - P_\text{FN}(\tau_m)) \geq prec )$}{
  \eIf{$(P_\text{FP}(\tau_l) - P_\text{FN}(\tau_l)) \times (P_\text{FP}(\tau_m) - P_\text{FN}(\tau_m)) \le 0$}{
  \texttt{//when $\tau_l < \hat\tau < \tau_m $ } \\
  $\tau_r \gets  \tau_m, Q_r \gets Q_m$ \\
  $P_\text{FP}(\tau_r) \gets P_\text{FP}(\tau_m), P_\text{FN}(\tau_r) \gets P_\text{FN}(\tau_m)$
  }{
  \texttt{//when $\tau_m < \hat\tau < \tau_r $ } \\
  $\tau_l \gets \tau_m, Q_l \gets Q_m$,\\
  $P_\text{FP}(\tau_l) \gets P_\text{FP}(\tau_m), P_\text{FN}(\tau_l) \gets P_\text{FN}(\tau_m)$ }

  $Q_m \gets \lfloor \frac{Q_l + Q_r}{2} \rfloor$  \\
  $\tau_m \gets \Percentile (Q_m) $ \\
  $P_\text{FP}(\tau_m), P_\text{FN}(\tau_m) = \FPRFNR (\boldmath{s}, \boldmath{y}, \tau_m)$ \\
  }
  $EER = \frac{P_\text{FP}(\tau_m) + P_\text{FN}(\tau_m) }{2}$ \\
\Return $EER$
\end{algorithm}
\end{fullwidth}

 \subsection{Database}

We used the publicly available PartialSpoof\footnote{https://zenodo.org/record/5766198} database to calculate EER. The PartialSpoof database \cite{zhang2021PartialSpoofDatabase} was generated by randomly substituting spoof (or bona fide) speech segments as bona fide (or spoof) from the same speaker. Bona fide and spoof segments were concatenated using the overlap-add method.
% And they labeled both concatenated parts and the spoofed nonspeech parts as \textit{spoof}, and bonafide nonspeech parts as \textit{bonafide}. 

\subsection{Configuration}
We measured the EER on the most powerful CM \cite{zhang2022partialspoof} on the PartialSpoof database when we wrote this paper. It utilized the self-supervised learning (SSL) model w2v2-large \cite{BaevskiZMA20-w2v2} as the front-end, gMLP \cite{Liu2021gmlp} as the back-end, P2SGrad-based mean squared error \cite{wang2021comparative} as the loss function, and Adam as the optimizer. It supports training using multiple resolutions or a single resolution. The finest resolution is at a frame level of 20 ms based on the configuration of the SSL model, and the coarsest segment-level resolution is 640 ms. We compared outputs extracted from branches of different resolutions trained at multiple resolutions and then discussed the relationship between them. We set $prec = 1e-5$ to estimate the range-based EER.

\subsection{Up-sampling and down-sampling predicted scores}
Models trained at fixed resolutions can usually only produce 
scores for uniform segments of the same resolution used during training. Then, we usually evaluate the performance using pre-segmented labels with the same resolution. If we want to measure the performance at different resolutions, we need to perform additional post-processing, like up-sampling or down-sampling, to convert predicted scores to the target measurement resolution. In this paper, as shown in the Hypothesis row of Fig.~\ref{fig:sco_gap_C2F}, (1) when up-sampling predicted scores to a fine-grained resolution, we duplicated each segment score following the relationship between the source resolution and fine-grained resolution,
 % we duplicated each segment score following the multiplicative relationship between the coarser-grained and fine-grained resolution; 
and (2) when down-sampling scores to a coarser-grained resolution, we aggregated adjacent segments by selecting their minimum\footnote{A segment with a lower score is more likely to be spoofed.} value.

% \begin{figure*}[!th]
% \centerline{\includegraphics[width=1.8\columnwidth]
% {Figures/score_upsample_downsample.pdf}}
% \caption{Procedure for upsampling and downsampling predicted scores.}
% \label{fig:up_down_sampling}
% \end{figure*}

\subsection{Results and discussion}
\label{sec:res}

\begin{table*}[!bth]
\caption{Range-based and point-based EER (\%) of multi-reso. CM in PartialSpoof.}
\label{tab:EER_PS}
\renewcommand\arraystretch{1.1}
\centering
\scriptsize
\resizebox{\linewidth}{!}{
\begin{tabular}{rcccccccccccccccc}
% \begin{tabular*}{\linewidth}{@{\extracolsep{\fill}} rcccccccccccccccc }
\toprule
    & \multicolumn{8}{c}{Development set}                                                                                         & \multicolumn{8}{c}{Evaluation set} \\
    \cmidrule(r){2-9} \cmidrule(l){10-17}
 Reso. of   & Range-    & \multicolumn{7}{c}{Point-based EER}                                                                   & Range-    & \multicolumn{7}{c}{Point-based EER}                                                                      \\
 \cmidrule(r){3-9}  \cmidrule{11-17} 
Training    & based EER & 10    & 20            & 40            & 80            & 160           & 320           & 640           & based EER & 10    & 20             & 40             & 80             & 160           & 320           & 640           \\
    \midrule
20  & 24.39     & 23.48 & \phantom{0}\textit{0.86} & \phantom{0}0.84  & \phantom{0}0.83 & \phantom{0}0.82 & \phantom{0}0.82  & 0.77          & 30.40     & 29.78 & \textit{12.84} & 11.94          & 10.52          & \phantom{0}8.42          & \phantom{0}5.96          & 4.06          \\
40  & 24.48     & 23.61 & 23.62         & \phantom{0}\textit{0.83} & \phantom{0}0.81          & \phantom{0}0.80          & \phantom{0}0.79          & 0.74          & 30.11     & 29.93 & 29.94          & \textit{11.94} & 10.51          & \phantom{0}8.43          & \phantom{0}5.98          & 4.10          \\
80  & 24.60     & 23.56 & 23.56         & 23.56         & \phantom{0}\textit{0.81} & \phantom{0}0.79          & \phantom{0}0.78          & 0.71          & 30.65     & 30.12 & 30.12          & 30.15          & \textit{10.92} & \phantom{0}8.70          & \phantom{0}6.14          & 4.15          \\
160 & 25.55     & 24.37 & 24.37         & 24.36         & 24.39         & \phantom{0}\textit{0.79} & \phantom{0}0.77          & 0.72          & 31.36     & 30.49 & 30.50          & 30.52          & 30.56          & \phantom{0}\textit{9.24} & \phantom{0}6.40          & 4.11          \\
320 & 30.03     & 28.99 & 28.99         & 28.99         & 29.02         & 29.09         &\phantom{0} \textit{0.75} & 0.69          & 33.91     & 33.39 & 33.38          & 33.41          & 33.45          & 33.48         & \phantom{0}\textit{6.34} & 3.97          \\
640 & 34.96     & 34.84 & 34.84         & 34.85         & 34.87         & 34.87         & 34.59         & \textit{2.15} & 37.38     & 37.53 & 37.53          & 37.54          & 37.56          & 37.56         & 37.54         & \textit{5.19}
\\
\bottomrule
\end{tabular}
}
\end{table*}

\begin{figure*}[!t]
% \centerline{\includegraphics[width=1.9\columnwidth]{Figures/gap.pdf}}
\centerline{\includegraphics[width=1.9\columnwidth]{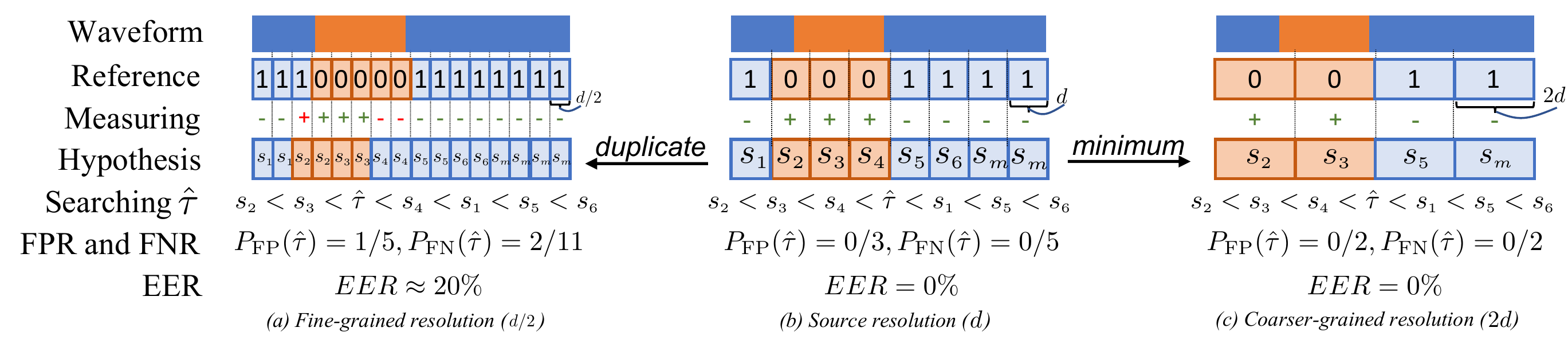}}
\vspace{-5mm}
\caption{An example for changing of score and error rate when up-sampling predicted score to fine-grained level (left) and down-sampling to coarser-grained level (right). [Given a threshold $\hat\tau$, the symbols ``+'' and ``-'' represent the predicted class as positive (spoof) and negative (bona fide) respectively. The color green indicates a correct prediction, while red represents a false prediction.]}
\label{fig:sco_gap_C2F}
\
\end{figure*}

Table \ref{tab:EER_PS} shows the results for models evaluated on the development and evaluation set of PartialSpoof separately. Each row has the same original predicted scores, and each column presents the measurement resolution. Thus, the point-based EER on the diagonal presents the training and measuring at the same resolution. The upper triangle of tables, from left to right, shows down-sampling predicted scores to coarser-grained resolution, and the lower triangle of those tables, from right to left, shows up-sampling predicted scores to fine-grained resolution.
% We analyze the results from two aspects in this section.

% \subsection{Relationship between point-based and range-based EER}
% \subsection{Analyses of Gap When Up-sampling}
From the point-based EER in Table \ref{tab:EER_PS}, we can notice that although each row has the same predicted score, different measurement resolutions can lead to significantly different performances. This is because the point-based references were defined from the pre-defined resolution. In the upper triangle (i.e., where the temporal resolution of the point-based reference is coarser than the temporal resolution of the training data), point-based EER may be an ``underestimation'' in terms of the spoof localization performance, since the reference becomes too coarse and does not reflect accurate boundary information as shown in Fig.~\ref{fig:sco_gap_C2F}. In such cases, although the error value becomes smaller, it just indicates that the task is easier and does not mean that spoof localization is more accurate. In general, errors must be interpreted with caution and in consideration of the temporal resolution used for the point-based EER.
On the other hand, when the temporal resolution of the point-based reference is finer than the temporal resolution of the training data (lower triangle of Table \ref{tab:EER_PS}), or a range-based reference is used (column ``Range-based EER'' of Table \ref{tab:EER_PS}), the error value will be naturally larger since the reference is more accurate and we can therefore account for errors at a finer level. Note that for the same row, even if the error is higher, it does not mean that the model is inaccurate -- they shared the same original predicted scores but were evaluated on references with different temporal resolutions.
% \textcolor{red}{And such evaluation can be easily influenced by the resolution of the references.}

Thus, for fair and proper evaluation of spoof localization models, we recommend using range-based EER, or point-based EER that uses references with a finer temporal resolution than that used during model training. In addition, when the training temporal resolution is unknown, the range-based EER would be a more appropriate choice.
% (we use 640 ms to 320 ms in as example)

% \subsection{Comparison on training resolution}
% Observing each column independently, we can see that the predicted score derived from the finest resolution (20 ms) performs better in most cases. This indicates that utilizing training data with fine-grained labels for spoof localization is the critical determinant for high-performing spoof localization.

\section{Conclusion}
\label{sec:conclusion}
In this paper, we first defined range-based EER for spoof localization and then adapted the binary search algorithm to estimate it. We finally utilized range-based EER and classical point-based EER to analyze the performance of spoof localization deeply and discussed the relationship between them. 
% We derived two recommendations for spoof localization from analyses: (1) For the training, we recommend using training data with fine-grained annotations. (2) 
For the measurement, we recommend using range-based EER, or point-based EER with unseen and finer temporal resolutions compared with the training resolution to more fairly and properly evaluate the performance of spoof localization.

\section{Acknowledgements}
This study is partially supported by the Japanese-French joint national VoicePersonae project supported by JST CREST (JPMJCR18A6, JPMJCR20D3), JPMJFS2136 and the ANR (ANR-18-JSTS-0001), MEXT KAKENHI Grants (21K17775, 21H04906, 21K11951), Japan, and Google AI for Japan program.

\bibliographystyle{IEEEtran}
\bibliography{main}

\newpage
\clearpage
\onecolumn

\begin{appendices}
\section{Appendix}
\subsection{Results on single-reso. CM.}

\begin{table*}[!bth]
\caption{Range-based and point-based EER (\%) of single-reso. CM in PartialSpoof.}
\label{tab:EER_PS_single}
\renewcommand\arraystretch{1.1}
\centering
\scriptsize
\resizebox{\linewidth}{!}{
\begin{tabular}{rcccccccccccccccc}
% \begin{tabular*}{\linewidth}{@{\extracolsep{\fill}} rcccccccccccccccc }
\toprule
    & \multicolumn{8}{c}{Development set}                                                                                         & \multicolumn{8}{c}{Evaluation set} \\
    \cmidrule(r){2-9} \cmidrule(l){10-17}
 Reso. of   & Range-    & \multicolumn{7}{c}{Point-based EER}                                                                   & Range-    & \multicolumn{7}{c}{Point-based EER}                                                                      \\
 \cmidrule(r){3-9}  \cmidrule{11-17} 
Training    & based EER & 10    & 20            & 40            & 80            & 160           & 320           & 640           & based EER & 10    & 20             & 40             & 80             & 160           & 320           & 640           \\
    \midrule
20  & 24.73     & 24.19           & \phantom{0}\textit{0.79} & \phantom{0}0.79          & \phantom{0}0.82          & \phantom{0}0.88          & \phantom{0}0.95          & \phantom{0}0.94          & 29.27     & 29.18           & \textit{10.24} & \phantom{0}9.64           & \phantom{0}8.70          & \phantom{0}7.27          & \phantom{0}5.50          & \phantom{0}4.10          \\
40  & 24.64     & 23.86           & 23.86         & \phantom{0}\textit{0.80} & \phantom{0}0.80          & \phantom{0}0.84          & \phantom{0}0.87          & \phantom{0}0.87          & 29.95     & 29.43           & 29.43          & \textit{10.53} & \phantom{0}9.38          & \phantom{0}7.71          & \phantom{0}5.72          & \phantom{0}4.17          \\
80  & 24.66     & 23.71           & 23.71         & 23.71         & \phantom{0}\textit{0.88} & \phantom{0}0.86          & \phantom{0}0.87          & \phantom{0}0.81          & 29.36     & 28.86           & 28.86          & 28.89          & \phantom{0}\textit{8.56} & \phantom{0}7.04          & \phantom{0}5.26          & \phantom{0}3.82          \\
160 & 25.68     & 24.70           & 24.70         & 24.69         & 24.72         & \phantom{0}\textit{0.94} & \phantom{0}0.88          & \phantom{0}0.85          & 29.83     & 29.05           & 29.06          & 29.09          & 29.13         & \phantom{0}\textit{6.94} & \phantom{0}5.12          & \phantom{0}3.71          \\
320 & 31.82     & 30.55           & 30.54         & 30.54         & 30.56         & 30.62         & \phantom{0}\textit{1.02} & \phantom{0}0.93          & 33.84     & 33.71           & 33.71          & 33.73          & 33.78         & 33.81         & \phantom{0}\textit{5.18} & \phantom{0}3.59          \\
640 & 35.48     & 35.11           & 35.10         & 35.10         & 35.14         & 35.13         & 34.78         & \phantom{0}\textit{2.87} & 38.36     & 38.08           & 38.08          & 38.09          & 38.11         & 38.12         & 38.07         & \phantom{0}\textit{6.12}   \\

\bottomrule
\end{tabular}
}
\end{table*}

\end{appendices}

\end{document}